\documentclass[10pt,a4paper,twocolumn,english,aps,prl,showpacs,floatfix,superscriptaddress,10pt]{revtex4-1}
\usepackage[T1]{fontenc}
\usepackage[latin9]{inputenc}
\usepackage{color}
\usepackage{babel}
\usepackage{geometry}
\usepackage{float}
\usepackage{amssymb}
\usepackage{graphicx}
\usepackage{breakurl}
\usepackage[unicode=true,
 bookmarks=true,bookmarksnumbered=false,bookmarksopen=false,
 breaklinks=false,pdfborder={0 0 0},backref=false,colorlinks=true]
 {hyperref}
\hypersetup{pdftitle={Evaluation of the spin diffusion length of AuW alloys using spin absorption experiments in the limit of large spin-orbit interactions.},
pdfauthor={Piotr Laczkowski}}

\makeatletter

\begin{document}

\title{Evaluation of the Spin Diffusion Length of $AuW$ Alloys Using Spin Absorption Experiments in the Limit of Large Spin-Orbit Interactions.}

\author{P. Laczkowski}
\affiliation{Unité Mixte de Physique CNRS-Thales, Univ. Paris-sud, Universit\'e Paris-Saclay 11, 91767 Palaiseau, France}

\author{H. Jaffrès}
\affiliation{Unité Mixte de Physique CNRS-Thales, Univ. Paris-sud, Universit\'e Paris-Saclay 11, 91767 Palaiseau, France}

\author{W. Savero-Torres }
\affiliation{Université Grenoble Alpes, INAC\textendash SP2M, F-38000 Grenoble, France}
\affiliation{CEA, INAC\textendash SP2M, F-38000 Grenoble, France}

\author{J.-C. Rojas-Sánchez }
\affiliation{Unité Mixte de Physique CNRS-Thales, Univ. Paris-sud, Universit\'e Paris-Saclay 11, 91767 Palaiseau, France}

\author{Y. Fu}
\affiliation{Université Grenoble Alpes, INAC\textendash SP2M, F-38000 Grenoble, France}
\affiliation{CEA, INAC\textendash SP2M, F-38000 Grenoble, France}

\author{N. Reyren}
\affiliation{Unité Mixte de Physique CNRS-Thales, Univ. Paris-sud, Universit\'e Paris-Saclay 11, 91767 Palaiseau, France}

\author{C. Deranlot}
\affiliation{Unité Mixte de Physique CNRS-Thales, Univ. Paris-sud, Universit\'e Paris-Saclay 11, 91767 Palaiseau, France}

\author{L. Notin}
\affiliation{Université Grenoble Alpes, INAC\textendash SP2M, F-38000 Grenoble, France}
\affiliation{CEA, INAC\textendash SP2M, F-38000 Grenoble, France}

\author{C. Beigné}
\affiliation{Université Grenoble Alpes, INAC\textendash SP2M, F-38000 Grenoble, France}
\affiliation{CEA, INAC\textendash SP2M, F-38000 Grenoble, France}

\author{J.-P. Attané}
\affiliation{Université Grenoble Alpes, INAC\textendash SP2M, F-38000 Grenoble, France}
\affiliation{CEA, INAC\textendash SP2M, F-38000 Grenoble, France}

\author{L. Vila}
\affiliation{Université Grenoble Alpes, INAC\textendash SP2M, F-38000 Grenoble, France}
\affiliation{CEA, INAC\textendash SP2M, F-38000 Grenoble, France}

\author{J.-M. George}
\affiliation{Unité Mixte de Physique CNRS-Thales, Univ. Paris-sud, Universit\'e Paris-Saclay 11, 91767 Palaiseau, France}

\author{A. Marty}
\affiliation{Université Grenoble Alpes, INAC\textendash SP2M, F-38000 Grenoble, France}
\affiliation{CEA, INAC\textendash SP2M, F-38000 Grenoble, France}

\date{\today}
\begin{abstract}
The knowledge of the spin diffusion length, $\lambda_{A}$, is a prerequisite for the estimation of the spin Hall angle of appropriate materials. We investigate the spin current absorption of materials with a short $\lambda_{A}$ using $AuW$ stripes inserted in $Cu$-based lateral spin-valves. Width variations of the $AuW$ stripe lead to drastic changes of the spin absorption which cannot be explained by a conventional analysis. We show that the spin-current polarization and the spin accumulation attenuation in $Cu$ in the vicinity of the spin absorber must be precisely taken into account for an accurate estimation of $\lambda_{A}$. We propose an analytical extension for the standard diffusion model of spin transport and spin absorption based on the existence of an effective spin diffusion length for $Cu$ being in direct contact with $AuW$. The calculations are supported by numerical investigations which allow to extract proper values of  $\lambda_{A}$.
\end{abstract}

\maketitle

\section{Introduction.}

Most of recent developments in spin-orbitronics have drawn an increased attention to the possibility of the efficient control of magnetization, the creation of a lateral charge current through bulk and/or surface properties involving strong spin-orbit interactions as well as spin-orbit assisted-scattering~\citep{Parkin_Science_2008,Miron_Nature_2011,Cubukcu_APL_2014,Parkin_NatNano_2015,Qiu_NatNano_2015}.
New synthesized materials exhibiting large Spin Hall Angles ($SHA$) \textit{via} their intrinsic and extrinsic properties~\citep{Sinova_PRL_2004,Sinitsyn_PRB_2007,Nagaosa_RMP_2010,Niimi_PRL_2011,Niimi_PRL_2012,Fujiwara_NatCom_2013,Niimi_PRB_2014,Liu_PRL_2011},
as well as carefully engineered Rashba-type interfaces~\citep{Sanchez_NatComm_2013,Borge_PRB_2014} and particular Fermi surface topologies~\citep{Moore_Nature_2010} have opened an access to an efficient spin-to-charge current conversion.
Moreover, the conversion ratio can be precisely tuned either by controlling the impurity level~\citep{Fert_JMMM_1981,Fert_PRL_2011,Grandhand_PRB_2010,Gradhand_SSP_2011}, gate voltage control~\citep{Caviglia_PRL_2010,He_APL_2012}, or even through magnetization control by a transverse spin absorption at spin-active magnetic insulator interfaces~\citep{Dejene_PRB_2015}.

Combined methods such as Ferromagnetic Resonance - Spin-Pumping~\citep{Ando_JAP_2010,Tserkovnyak_RMP_2005,mosendz_PRB_2010}, Spin Torque-Ferromagnetic Resonance~\citep{Liu_PRL_2011}, second harmonic Hall effect measurements \citep{Garello_NatureNano_2013} or Lateral Spin Valves ($LSV$)~\citep{Kimura_PRB_2005} give access to the $SHA$. However, concerning \textit{e.~g.} Pt, the reported values are spread over one order of magnitude~\citep{Rojas_PRL_2014}. Moreover, in order to accurately estimate the $SHA$ or the intrinsic spin Hall conductivity of a given material, its spin diffusion length ($SDL$) must be known. This can be achieved either by comparative non-local spin signal measurements (spin sink experiments) or by the examination of the thickness dependence of combined spin Hall effect ($SHE$) and spin-pumping experiments~\citep{Rojas_PRL_2014,Nakayama_Magn_2010,Azevedo_PRB_2011,Feng_PRB_2012}. However, at the nanometer scale, the latter method appears to be really challenging due to possible lack of film continuity for small thicknesses which are necessary for characterization of a short $SDL$. This precludes the use of this technique for thinnest layers. In contrast, $LSV$ designed for non-local spin-sink experiments provide an alternative which has already been proven to be efficient for short $SDL$ materials~\citep{Kimura_PRB_2005,Vila_PRL_2007}. In order to extract the $SDL$, simulations by Finite Element Method ($FEM$) is a pertinent approach~\citep{Niimi_PRL_2012}. Nevertheless, the exact knowledge of spin-current pathways in the limit of a short $\lambda_A$ requires a heavy mesh density and large computation power in the case of lateral inhomogeneous structures displayed here.

\begin{figure}
\centering{}\includegraphics[height=4.5cm]{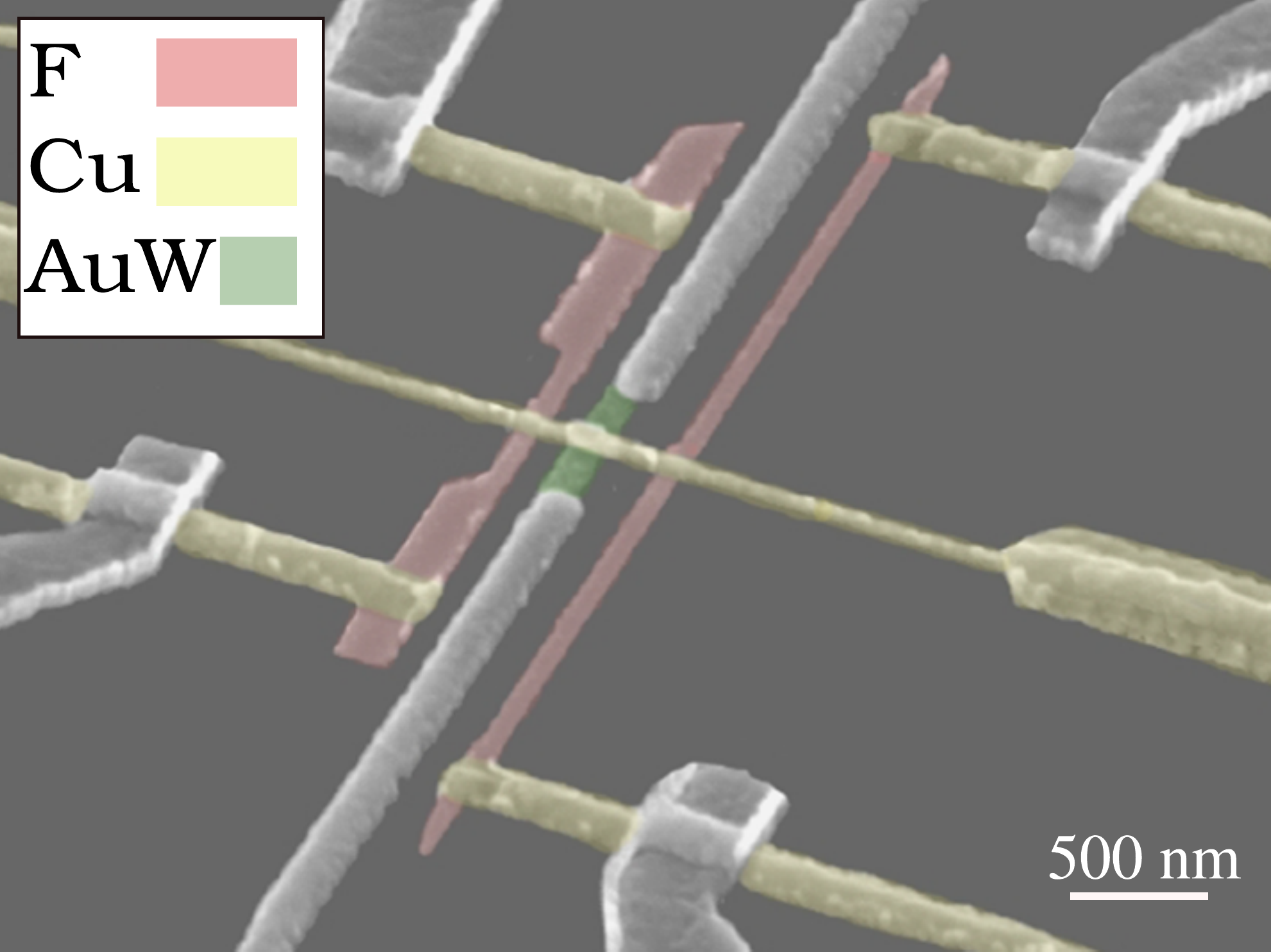}\caption{\textit{\label{fig:SEM_device}Scanning electron microscope image of a typical lateral spin valve nano-device fabricated using multi-angle evaporation technique with inserted $AuW$ spin absorber nano-wire }\citep{Laczkowski_APEX_2011}.}
\end{figure}

In this paper we present a series of experimental data from $LSV$ with the inserted metallic $AuW$ absorber exhibiting an extrinsic spin-Hall effect~\citep{Laczkowski_APL_2014} as well as a refined spin-absorption data analysis. In the present work the concentration of $W$ reaches 13.8\% in an the $Au$ host. The improved analysis is based on an extended 1-dimensional ($1D$) spin absorption model taking into account the varying spin-current polarization and spin accumulation profiles along the spin-absorber of the width $w_{A}$. We demonstrate that for the short $SDL$ of $AuW$, more exactly when $w_{A}$ becomes comparable to the effective $SDL$ of the non-magnetic channel ($N=Cu$), the lateral spread of both spin-current polarization and spin-accumulation along the spin-flow direction needs to be considered beyond the standard point-contact model. We show that these exact profile shapes become a relevant parameter when the spin-resistance of the spin-absorber $R_{A}$ becomes smaller than a certain threshold value involving the spin-resistance of the non-magnetic channel $R_{N}$ itself. We show that it can lead to significant errors in the evaluation of $\lambda_{A}$ reaching more than 90\% in the present case. We then propose a new criterium for the validity of the standard point-contact model, demonstrating that the situation is generally more complex. That means that a significant correction may be required even in the limit of a long $SDL$ of the non-magnetic material $\lambda_{N}$. We develop an extended model in the limit of full interface transparency \textit{i.e.} neglecting any interface resistances, taking into account a necessary renormalization of $\lambda_{N}$ nearby the contact region. We then extract a robust value of $\lambda_{A}=1.25~$nm for $AuW_{13.8\%}$ alloys independent of the contact width $w_{A}$.

\section{Experimental Details.}

In our experiments we consider metallic $LSV$ with transparent interfaces. The transparent interfaces are inferred from a 4-point measurement of each interface resistance, being of the order of $1 f{\Omega}.m^{2}$ or smaller. Fig.~\ref{fig:SEM_device} represents a scanning electron microscope image of a typical $LSV$ device fabricated using the multi-angle nano-fabrication technique~\citep{Laczkowski_APEX_2011}, where the red color represents the ferromagnetic ($F$) injector made of $Py$, the yellow one the nonmagnetic channel made of $Cu$ and the green one the spin-sink or spin-absorber ($A$) material. First, the middle $AuW_{13.8\%}$ wire is deposited on a $SiO_{2}$ substrate using sputtering and lift-off technique, followed by the nanofabrication of a $Py/Cu$ $LSV$. In-between these two steps the middle wire surface is cleaned using Ar ion-milling. The $Py$, $Cu$ and $AuW$ nanowires are respectively  $20\,nm$, $77\,nm$ and $30\,nm$ thick ($t$ stands for the thickness). Their width is fixed to $50\,nm$ with the exception of the $AuW_{13.8\%}$ nanowire which width was varied in a series of samples corresponding to $w_{A}=45,\,95,\,195\,nm$ in order to study the spin absorption. The distance separating the ferromagnetic electrodes is $L=600\;nm$ (from the center-to the center), whereas the spin-sink is placed exactly in the center. The device geometry as well as characteristic material resistivities, spin-current polarization and $SDL$ extracted from complementary measurements are given in Table $I$. The $SDL$ of $AuW$ given at $1.25$~nm corresponds to the value extracted from the following extended analysis.

\begin{center}
\begin{table}[H]
\begin{centering}
\begin{tabular}{cccccc}
$material$ & $w\,[nm]$ & $t\,[nm]$ & $\rho\,[\Omega.nm]$ & $P_{eff}$ & $\lambda\,[nm]$\tabularnewline
\hline
\hline
$F$ & $50$ & $15$ & $110$ & $0.35$ & $5.5$\tabularnewline
$N$ & $50$ & $77$ & $77$ &  & $380$\tabularnewline
$A$ & $50$ & $30$ & $1054$ &  & $1.25$\tabularnewline
\hline
\end{tabular}
\par\end{centering}
\caption{Table summarizing width, thickness, resistivity and effective spin-polarization of materials used in the lateral spin-valves devices.}
\end{table}
\par\end{center}

\begin{figure}
\begin{centering}
\includegraphics[height=13cm]{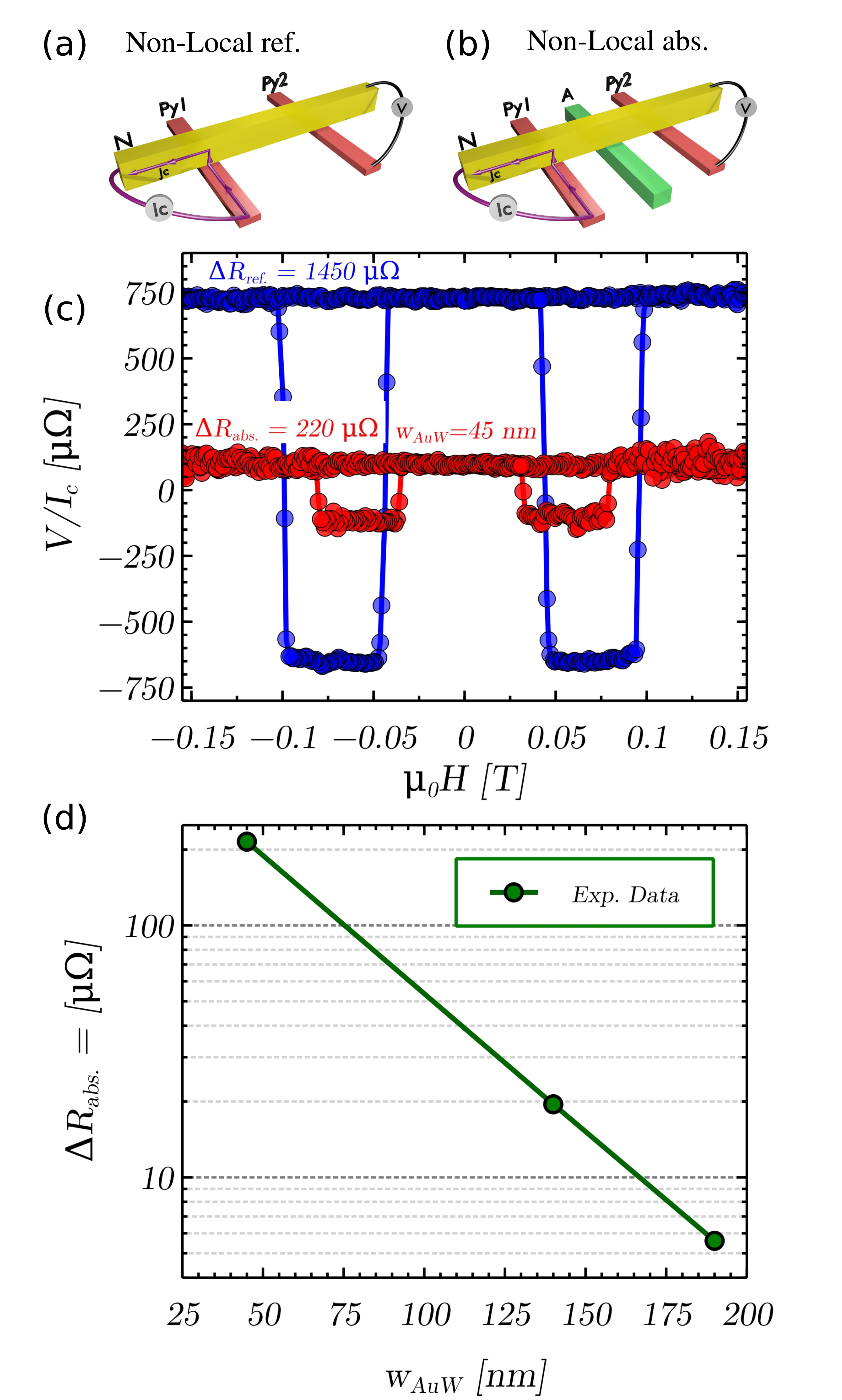}
\par\end{centering}

\caption{\textit{\label{fig:Experiments}Schematic representations of spin-sink experiment for (a) reference (without $AuW$ wire) and (b) absorption (with inserted $AuW$ wire) nano-devices measured in the non-local probe configuration; (c) Non-local measurements of reference (blue) and absorption (red) devices recorded at $T=9\,K$ for ferromagnetic electrode separation of $L=600\,nm$ (from the center of $Py1$ to the center of $Py2$) and $w_{A}=45\,nm$; (d) Amplitude of the spin signal of the absorption device in the non-local probe configuration as a function of the $AuW$ nanowire width.}}
\end{figure}

In these experiments, we follow the same protocol as for the $SDL$ evaluation presented in our previous work~\citep{Laczkowski_PhD_2012,Laczkowski_APL_2014}. Non-local measurements have been performed for two types of lateral nanodevices: the regular non-local device {[}Fig.~\ref{fig:Experiments}(a){]} without spin-sink used as the reference and the non-local device containing a $AuW_{13.8\%}$ wire inserted in-between the two ferromagnetic injectors {[}Fig.\ref{fig:Experiments}(b){]}. As a result, clear spin-signals were observed for both: the reference (in blue) and absorption (in red) devices. Their amplitudes were measured to be $\Delta R_{ref}=1.45\,m\Omega$ and $\Delta R_{abs}=0.22\,m\Omega$ for $w{}_{A}=45\,nm$~{[}Fig. \ref{fig:Experiments}(c){]}. The drop of the signal for the absorption device compared to the reference device indicates an efficient absorption of spin accumulation by the $AuW$ wire and a strong reduction of the spin-current reducing thus the overall magnetoresistance ($MR$), as expected. Also, the amplitude of the spin-signal was studied as a function of the $AuW$ nano-wire width, yielding $\Delta R_{abs}=220\,\mu\Omega,\,19.5\,\mu\Omega,\,5.6\,\mu\Omega$
for $w_{A}=45,\,140,195\,nm$ respectively. One can observe a clear decay of $\Delta R_{abs}$ when increasing the absorber width $w{}_{A}$~{[}Fig.~\ref{fig:Experiments}(d){]}. We define $\eta$ as the ratio between the $MR$ measured with and without the spin-absorber which manifests the efficiency of the spin-absorption \textit{i.e.} $\eta=\Delta R_{abs}/\Delta R_{ref}$.

\section{Standard Analysis.}

By comparing $\Delta R_{abs}$ with $\Delta R_{ref}$ and using the resistivity of the inserted $AuW$ wire $\rho_{A}=1040\,\Omega~nm$~\citep{Laczkowski_PhD_2012}, one can give a first rough evaluation of its spin diffusion length. For this purpose, the commonly used $1D$ point-contact model~\citep{Niimi_PRL_2011,Niimi_PRL_2012} is employed~\footnote{The $1D$ model used in these references is an advanced version of the one developed in~\citep{Kimura_PRB_2005}. However, a factor of two in the extracted $SDL$ can be systematically found between these two expressions, in the way that $\lambda_{A}^{Kimura\,et\,al.}=2\times\lambda_{A}^{Niimi\,et\,al.}$.}. As a rule, we define the spin resistances for the ferromagnets, the non-magnet and the spin absorber respectively as: $R_{F}^*=\frac{\rho_{F}\lambda_{F}}{(1-P_F^{2})w_{F}w_{N}}$, $R_{N}=\frac{\rho_{N}\lambda_{N}}{w_{N}t_{N}}$, $R_{A}=\frac{\rho_{A}\lambda_{A}}{w_{A}w_{N}}$. $P_F$ represents the bulk spin-polarization of $Py$ injectors. The values of $\lambda_{A}$ extracted from such conventional analysis \textit{vs.} $w_{A}$ are summarized in Table~\ref{tab:Table}:

\begin{center}
\begin{table}[H]
\begin{centering}
\begin{tabular}{cccc}
$w_{A}\,[nm]$ & $45$ & $140$ & $195$\tabularnewline
\hline
\hline
$\lambda_{A}\,[nm]$ & $1.32$ & $0.30$ & $0.10$\tabularnewline
\end{tabular}
\par\end{centering}
\caption{\textit{\label{tab:Table}The spin diffusion length of the }$AuW_{13.8\%}$\textit{nanowire $\lambda_{A}$ for three different $w_{A}$ estimated by using a $1D$ point-contact spin absorption model.}}
\end{table}
\par\end{center}

\begin{figure}
\begin{centering}
\includegraphics[width=8cm]{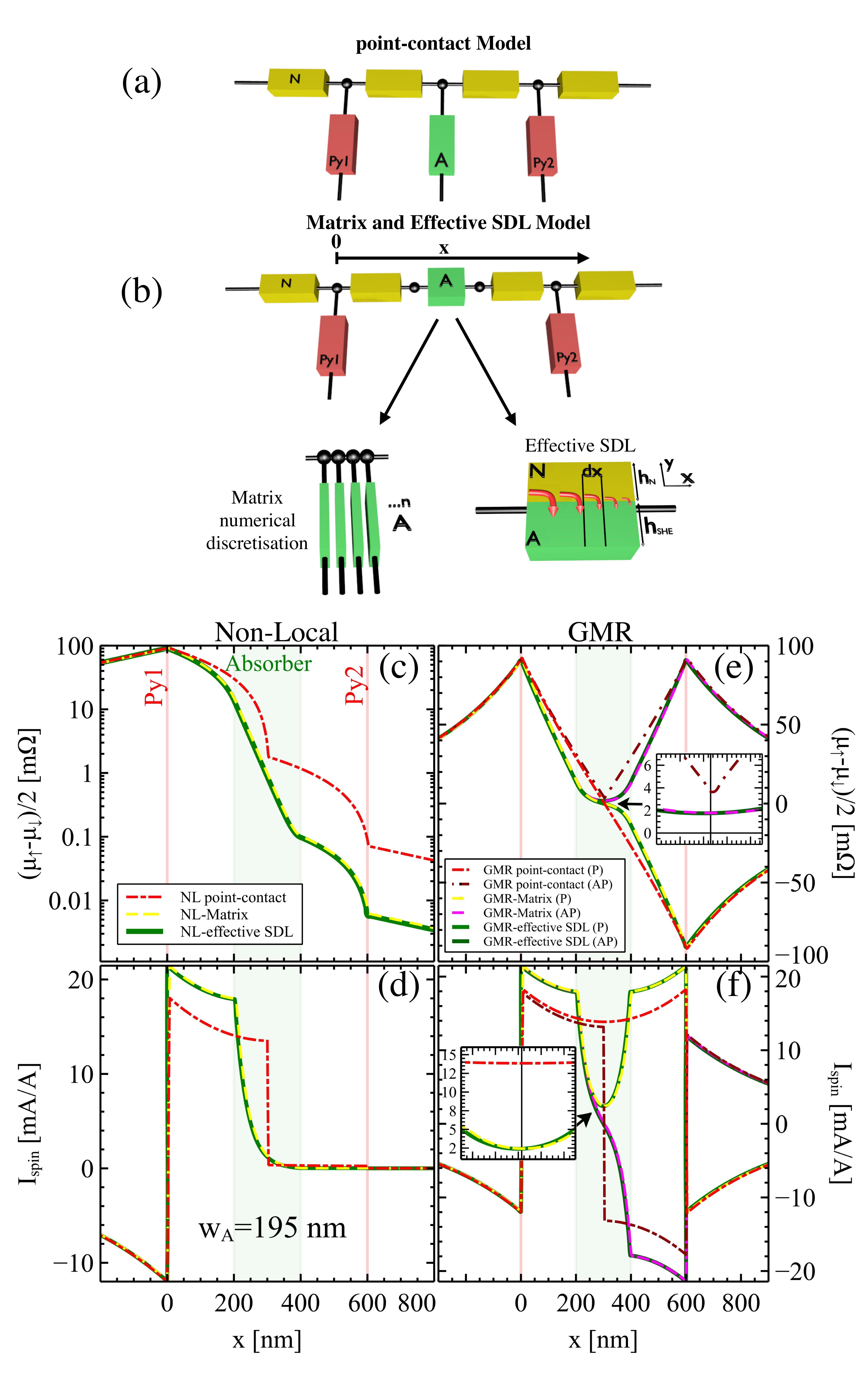}
\par\end{centering}

\caption{\textit{\label{fig:Models} Schematic representation (in the frame of the spin resistor approach) of (a) the point-contact and (b) the effective $SDL$ with two SHE resistor approaches represented at the bottom. Profiles of the spin accumulation (top) and the current spin polarization (bottom) for the point-contact (dash-dot), the transfer Matrix (dash) and the effective $SDL$ (solid line) models for (c-d) Non-Local and (e-f) Giant Magneto-Resistance ($GMR$) measurement configurations. Notations $'AP'$ and $'P'$ stand for the anti-parallel and parallel magnetic configurations of the Py injector/detector. The position of the injector ($Py1$) and the detector ($Py2$) as well as the spin absorber (Absorber) are identified by vertical lines.}}
\end{figure}

Contrary to what might be expected, different values of \textit{$\lambda_{A}$} are deduced \textit{vs.} $w_{A}$ for the same $W$ concentration. In particular, for the larger $AuW$ wire an extremely short $\lambda_{A}$ of about $0.1\,nm$ is found. This has no real physical meaning except indicating a larger spin absorption estimated this way. What is the main reason for that? As the width of the spin-sink increases, the assumption of a point-contact spin-sink (its schematic representation using the spin-resistor model~\citep{SaveroTorres_Xresistor_2015} is shown on Fig.~\ref{fig:Models}(a)) ceases to be valid due to the spatial variation of spin-current and spin-accumulation profiles in $N$. Indeed, the apparent decrease of \textit{$\lambda_{A}$} \textit{vs.} $w_{A}$ can be understood when one considers that the rate of spin-absorption is strongly inhomogeneous along the contact. This gives rise, in average, to different apparent values of $SDL$ in $A$, instead of the real physical ones. Moreover, we will show that a refined analysis is necessary well before the intuitive condition $w_{A}\gtrsim \lambda_N$ is reached because of a shorter effective $SDL$ in that region. We refer to the standard approach and formula given in Ref.~\citep{Kimura_PRB_2005} as the point-contact model.

Our refined effective $SDL$ approach is illustrated in Fig.~\ref{fig:Models}(b) and its validity will be checked numerically. It originates from the Spin-Resistor scheme described in details elsewhere~\citep{SaveroTorres_Xresistor_2015} and is based on the important following fundamentals:

\textit{i}) The spin-absorption in the spin-sink, at the level of the $Cu/AuW$ interface, originates from spin-diffusion/relaxation
processes along the direction normal to the junction. This is the general way to consider the spin-current dissipation in the standard Valet-Fert approach~\citep{ValerFert_PRB_1993}, extended in the present case to $LSV$ structures. The important physical parameters involved are the spin-flip resistance (with calligraphic notations) which, for thin layer, differs from the spin-resistance itself according to the general formula $\mathcal{R}_s=\rho [\lambda_s]^2/V_{sf}$. $\lambda_s$ is the $SDL$ and $V_{sf}$ is the total spin-flip volume in each media, $\mathcal{R}_N$ for the channel and $\mathcal{R}_A$ for the spin-sink. In the particular geometry of a long $\lambda_{N}$ and short $\lambda_{A}$ (compared to thickness), $\mathcal{R}_N=\rho_N [\lambda_N]^2/(S_A^* t_N)$ and $\mathcal{R}_A=\rho_A \lambda_A/S_A^*$ where $S_A^*$ is the effective contact (spin-flip) area. In other type of spin-injection experiments, the enhancement factor $\lambda_N/t_N$ appearing above in the expression of $\mathcal{R}_N$ is also responsible for the increase of the spin-signal at oxide-semiconductor interfaces (the Hanle effect) in the limit of a long spin-flip time $\tau_{sf}\propto [\lambda_N]^2$~\citep{Tran_PRL_2009} as well as in graphene-based devices for spin-amplification as we pointed out in Ref.~\citep{Laczkowski_PRB_2012}. The case $\mathcal{R}_N\ll \mathcal{R}_A$ corresponds to a majority spin-relaxation in $N$ whereas $\mathcal{R}_A\ll \mathcal{R}_N$ corresponds to a majority spin-relaxation in $A$. One may expect a strong correction to occur in the analysis when the condition $\mathcal{R}_A\ll \mathcal{R}_N$ is realized, \textit{i.e.} when spins have a large access to the spin-sink material. This condition matches with the condition $\sqrt{\frac{\rho_A}{\rho_N} \lambda_A t_N}\ll \lambda_N$. This condition will be recovered below from pure analytical analysis.

\textit{ii}) The amount of spin diffusion/relaxation process (spin-current exchange at interfaces) depends on the possible spin-accumulation profile along the in-plane spin-flow direction within the non-magnetic channel. This particular feature is not addressed in the standard point-contact model in lateral devices. In order to address this issue, an effective $SDL$ in the contact region should be defined (this work). The alternative consists in a numerical procedure \textit{via} discretization of the spin-sink region into several elements and performed in the frame of the point-contact model adapted to these multielements. The solution can be implemented using the transfer matrix method. The two methods, effective $SDL$ and matrix numerical discretization, will be shown to give identical results.

\textit{iii}) From the general arguments developed above, once one admits that the amount of spin-absorption linearly depends on the integrated spin-accumulation profile, one can expect an up-renormalization of the spin-absorption by a certain factor.

\textit{iv}) The method will be proven to be valid if one ignores any possible lateral spin-current in the spin-absorber itself
(case of very short $\lambda_{A}$). This particular assumption, often justified in case of a strong spin-absorber, gives an upper bound of the normalization to consider for the determination of the effective $SDL$ in $N$, $\lambda_N^*$. Indeed, the opposite limit of a very long $\lambda_{A}$ compared to its width $w_A$ undoubtedly leads to the condition $w_A\ll \lambda_N^*$ ruling out any necessary renormalization.

\textit{iv}) In this work, we disregard possible discretization effects in the regions of the outward ferromagnetic $F$ injectors.
However, a width of $F$ of the order of the effective $SDL$ will also require similar treatment as for the spin-absorber. We have checked, that the improved method adapted also to the $F$ injectors leads only to minor changes in the spin-signals for the present study.

\section{Refined Analysis: the effective SDL model.}

First, we consider the analytical problem of a spin-sink of width $w_A$ and $\lambda_A$ in contact with a non-magnetic channel of thickness $t_N$ and $\lambda_N$. This part of the system may be viewed as a unique media (non-magnetic channel) described by an effective SDL~{[}Fig. \ref{fig:Models}(a){]}. To see this, one defines the spin-accumulation in $N$ by $\Delta \mu=(\mu_\uparrow-\mu_\downarrow)/2$, spin-current by $J_s=(J_\uparrow-J_\downarrow)$ and describes the spin-current continuity equation, $\nabla J_s=-\Delta \mu/[\rho_N (\lambda_N)^2]$ as follows:

\begin{eqnarray}
-\frac{\Delta \mu}{\rho_N(\lambda_{N})^{2}}=\frac{\partial J_{s,x}}{\partial x}+\frac{J_{s,y}}{t_{N}}\nonumber
\end{eqnarray}
where we have assumed a thickness $t_N$ well smaller than $\lambda_N$. $J_{s,y}$ is the vertical spin current density at the $N/A$
interface. The spin current density along the lateral direction, $J_{s,x}$ is derived from the spin-accumulation profile according to:
\begin{eqnarray}
J_{s,x}=-\frac{1}{\rho_N }\frac{\partial \Delta \mu}{\partial x}\nonumber
\end{eqnarray}

As for the vertical spin current density at the interface, it can be obtained through the following relationship while neglecting lateral spin current inside the spin-sink:
\begin{eqnarray}
J_{s,y}=\frac{\Delta \mu}{\rho_{A}\lambda_{A}}\tanh(\frac{t_{A}}{\lambda_{A}})\nonumber
\end{eqnarray}
which describes a total spin-current dissipation in the spin-sink by spin-flip processes. We have also neglected any possible interface resistance which effect would have been to make spin-accumulation discontinuous at either side of the interface. We derive:

\begin{eqnarray}
\frac{\partial \Delta \mu}{\partial x^{2}}=\frac{\Delta \mu}{[\lambda_{N}^*]^{2}}\\
\frac{1}{(\lambda_{N}^*)^{2}}=\frac{1}{(\lambda_N)^2}+\frac{\rho_N}{\rho_A}\frac{\tanh(\frac{t_A}{\lambda_A})}{\lambda_A t_N}\nonumber \\
\frac{1}{(\lambda_{N}^*)^{2}}=\frac{1}{[\lambda_N]^2} \left(1+\frac{\mathcal{R}_N}{\mathcal{R}_A} \tanh(\frac{t_A}{\lambda_A}) \right)
\end{eqnarray}
which is the equation we are looking for. The effective SDL in the spin-sink region equals $\lambda_{N}^*=\frac{\lambda_N}{\sqrt{1+\frac{\mathcal{R}_N} {\mathcal{R}_A} \tanh(\frac{t_A}{\lambda_A})} }$ which expresses a reduction of the spin-diffusion length as $\mathcal{R}_A\ll \mathcal{R}_N$ when $\lambda_N\ll \sqrt{\frac{\rho_A}{\rho_N} \lambda_A t_N}$. This corresponds to the condition derived from simple arguments, as presented in part III. It is then expected that a renormalization will become necessary for contact width $w_A\gg \lambda_{N}^*$ with $\lambda_{N}^* \simeq 35$~nm in the present case. We are now going to demonstrate how the varying spin-accumulation and spin-current profiles in the non-magnetic channel $N$ may affect (enhance) the spin-absorption process and how they are responsible for the drop of the spin-signal $MR$.

\subsection{Spin-accumulation and spin-current profiles}

We first discuss the issues of spin-accumulation and spin-current polarization profiles along the $Cu$ channel. Concerning the numerical procedure, we discretize the $AuW$ contact into a collection of at least 9 same elements of equivalent size over the whole experimental series ($45$ to $195\,~nm$ width), sufficient to converge towards a common robust value for the $SDL$. Each discrete element is then of a maximum lateral size of $20\,nm$ for the largest wire which will correspond to about $\lambda_{N}^*/2\simeq 20$~nm.

We have compared both the spin-accumulation and spin-current polarization profiles for the cases of a non-local {[}Fig. \ref{fig:Models}(b){]} and $GMR$
(two-contact) probe configurations \citep{Laczkowski_PRB_2012} with $w_{A}=195\,nm$ \footnote{Note that the Non-Local configuration corresponds to the injection of current at the left ferromagnetic wire and the absorption at infinity whereas in the $GMR$ probe configuration the current is injected at the left ferromagnetic wire and absorbed at the right ferromagnetic wire.}. These are displayed on ~Fig.~\ref{fig:Models}(c-f). Both calculations, effective $SDL$ and matrix numerical discretization, lead to the exact same profiles proving the validity of the effective $SDL$ method. We also checked that this match is valid for every geometry considered. In particular, in the non-local geometry these calculations emphasize a decrease of 70\% of the level of spin accumulation at the center of the spin-sink {[}Fig. \ref{fig:Models}(c){]} compared to the standard approach together with a strong increase of the spin-current by 93\% {[}Fig. \ref{fig:Models}(d){]} at the same point due to the shorter effective $SDL$. However, in a standard 2-point $GMR$ configuration the spin-accumulation decreases by about 50\% in the $AP$ state {[}inset Fig. \ref{fig:Models}(e){]} whereas the spin-current polarization drops by 90\% {[}inset of Fig. \ref{fig:Models}(f){]} in the $PA$ state. This should be correlated to a total decrease of the spin-signal $\Delta R=2P^{\uparrow \uparrow} \Delta \mu^{\uparrow \downarrow}$~\citep{Laczkowski_PRB_2012} ($\uparrow \uparrow, \uparrow \downarrow$ represent the $PA$ and $AP$ states respectively~\citep{Jaffres_PRB_2010}) by about 95\% compared to the conventional point-contact analysis. The conclusion is that the point-contact approach obviously fails in the determination of both the spin-current polarization and spin-accumulation profiles.

\begin{figure}
\begin{centering}
\includegraphics[height=12cm]{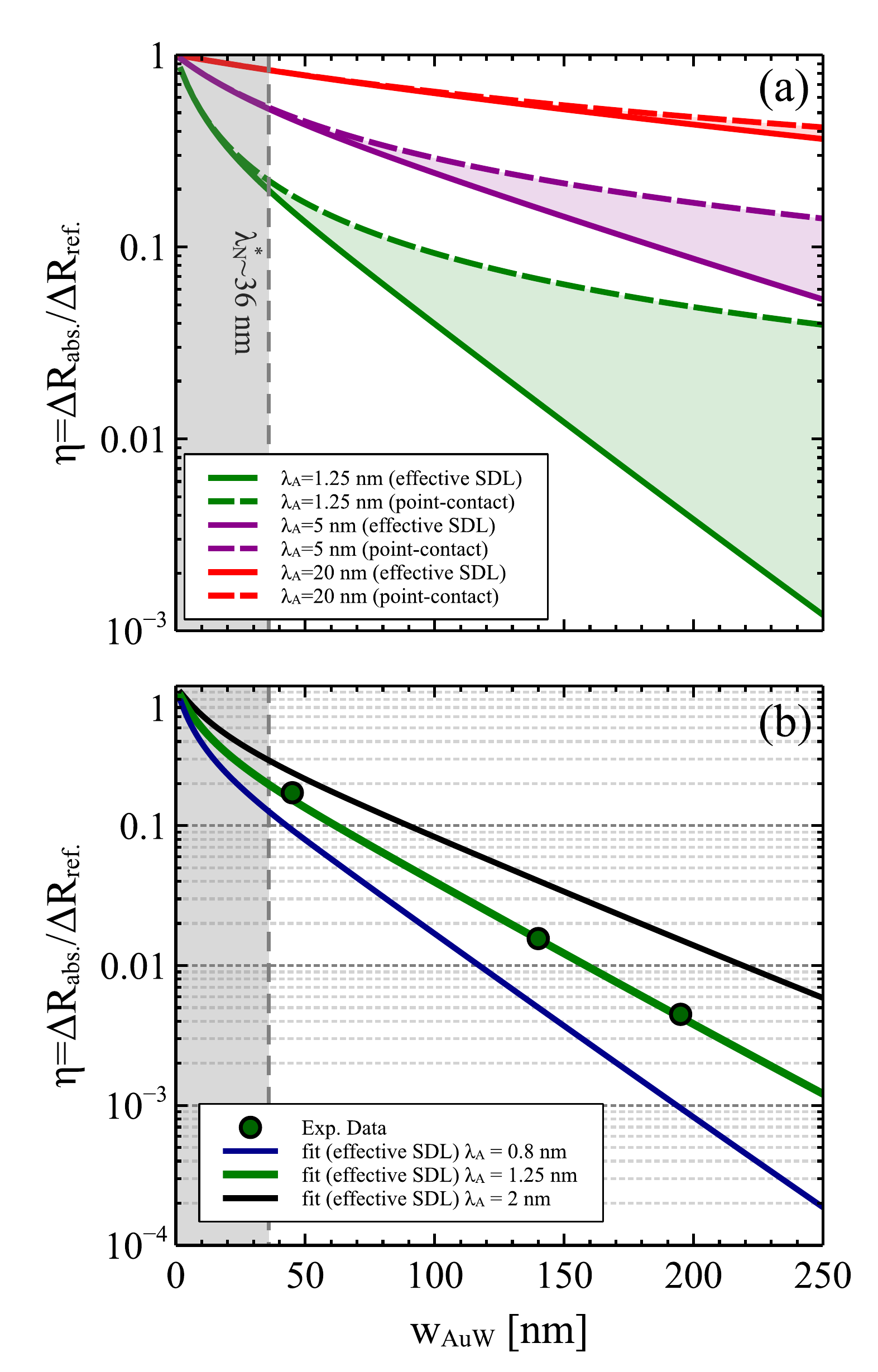}
\par\end{centering}

\caption{\textit{\label{fig:Fits}(a) Normalized spin signal amplitude $\eta$ as a function of the width of the $AuW$ nano-wire $w_{A}$ represented for a case of three different spin diffusion lengths $\lambda_{A}$. Dashed line stands for point-contact while solid line stands for Resistor model. (b) $AuW$ spin diffusion length fits using Resistor model (solid lines) to the experimental data (green points) plotted for three different cases. Grey region with dashed line indicates the limit where }$w_{A}<\lambda_{N}^*$. All materials characteristic parameters used in these calculations are summarized in the Table I.}
\end{figure}

\subsection{Spin-signals and Magnetoresistance}

In order to get more insight into the limits of validity of the point-contact model, we have compared the normalized spin signal $\eta=\frac{\Delta R_{abs.}}{\Delta R_{ref.}}$ \textit{vs.} the absorber width $w_{A}$ for both the point-contact and the effective $SDL$ models. Fig.~\ref{fig:Fits}(a) displays, in each case, the calculated spin-resistance \textit{vs.} $w_{A}$ for three different $SDLs$ denoted by different colors. This plot highlights important differences between the point-contact and effective $SDL$ especially for the case of short $\lambda_{A}$ and in the limit where {$w_{A}\geq\lambda_{N}^*$.} For the larger nanowires \textit{i.e.} $w_{A}=200\,nm$ the difference in the $\eta$ can reach 95\% leading to an incorrect value of $\lambda_{A}\sim 0.1\,nm$. One may notice significant differences for all $\lambda_{N}<2\,\mu m$ while keeping $R_{N}$ constant. These differences are however not so pronounced if one considers a longer $SDL$ in the absorber $\lambda_{A}\sim12.5\,nm$ when using $\lambda_{N}\sim 1.5\,\mu m$ and keeping $R_{N}$ constant. We come to the conclusion that the geometrical renormalization of the spin signal leading to an increase of the spin absorption is necessary in the case of a short $\lambda_{A}$ ($1.25\,nm$ in the present case) even if $\lambda_{N}$ is large up to $1.5\,\mu m$.

Fig.~\ref{fig:Fits}(b) displays the fit of the experimental data of $\eta$ as a function of $w_{A}$, by using the effective $SDL$ (green line). The black and blue lines represent the case of different $\lambda_{A}$ using the same model. They are displayed to demonstrate the high precision we get by using the refined analysis on $\lambda_{A}$. We highlight that a single and robust value of $\lambda_{A}$ is obtained for three experimental data-points, as summarized in Table ~\ref{tab:Table} and that the extrapolation to $w_{A}=0\,nm$ allows to recover the spin-signal of the reference device $\Delta R_{ref}=1.45\,m\Omega$.

\subsection{Trends for the geometrical renormalization}

What are the main trends of the effective $SDL$ model?  One may define the effective resistance to spin-flip in the absorber region as $\mathcal{R}_A^*=[(\mathcal{R}_N)^{-1}+(\mathcal{R}_A)^{-1}]^{-1}=\rho_N [\lambda_N^*]^2/(S_A^* t_N)$ and the ratio $r=\mathcal{R}_A/\mathcal{R}_N$. The effective $SDL$ will have a large effect for $r\ll 1$. We note respectively the two series-resistances $\mathcal{D}=R_F^*+R_{ch}+\mathcal{R_A}$ and $\mathcal{D}^*=R_F^*+R_{ch}+\mathcal{R_A^*}$ where $R_{ch}=\rho_N (L-w_a)/2$ is the channel resistance in series between $F$ and $A$. We recall that $\eta=\frac{\Delta R_{abs.}}{\Delta R_{ref.}}$ is the ratio between the spin-signals for absorption and reference devices respectively. The effective spin-flip surface area in $AuW$ is proportional to $w_A$ in the point-contact model and $2\lambda_N^*$ in the real situation (effective $SDL$). We propose the following rules for the determination of $\lambda_A$ in the limit of a small resistance-to-spin-flip $\mathcal{R}_A\ll \mathcal{R}_N$ (or equivalently $r\ll 1$). This limit gives also $\mathcal{D}^*=\mathcal{D}$.

\textit{i}) Renormalization of the standard $1D$ point-contact model is necessary when $w_A \gtrsim \lambda_{N}^*=\frac{\lambda_N}{\sqrt{1+\frac{{R}_N}{{R}_A}}}$, in the region of the spin-absorber $A$.

\textit{ii}) The polarisation of the spin-current $\mathcal{P}^{\uparrow \uparrow}$ in the $PA$ state at the center of the spin-absorber approaches $\mathcal{P}^{\uparrow \uparrow}=P_F \frac{R_F^*}{\mathcal{D}^*}\exp{[-(L-w_a)/(2\lambda_N)]}\exp{[-w_a/(2\lambda_N^*)]}$. The point-contact model gives $\mathcal{P}^{\uparrow \uparrow}=P_F \frac{R_F^*}{\mathcal{D}}\exp{[-(L-w_a)/(2\lambda_N)]}\exp{[-w_a/(2\lambda_N)]}$.

\vspace{0.1in}

\textit{iii}) The level of spin-accumulation $\Delta \mu^{\uparrow \downarrow}$ in the $AP$ state at the center of the spin-absorber approaches $\Delta \mu^{\uparrow \downarrow}=P_F \frac{R_F^* \mathcal{R}_A^*}{\mathcal{D}^*}\exp{[-(L-w_a)/(2\lambda_N)]}\exp{[-w_a/(2\lambda_N^*)]}$. The limit of the point-contact model gives $\Delta \mu^{\uparrow \downarrow}=P_F \frac{R_F^* \mathcal{R}_A}{\mathcal{D}}\exp{[-(L-w_a)/(2\lambda_N)]}\exp{[-w_a/(2\lambda_N)]}$.

\vspace{0.1in}

\textit{iv}) The effective surface of spin absorption is $S_A^*=w_N w_A$ and $S_A^*=2 w_N \lambda_N^*$ for the point-contact and the effective $SDL$ models respectively.

\vspace{0.1in}

\textit{v}) The $MR$ writes $\Delta R = R^{\uparrow \downarrow}-R^{\uparrow \uparrow}= 2\mathcal{P}^{\uparrow \uparrow} \Delta \mu^{\uparrow \downarrow}$. From the two upper relationships, we have $\frac{\eta_{eff}}{\eta_{pc}}=\frac{\mathcal{R}_A^*}{\mathcal{R}_A}\frac{\exp{[w_a/(2\lambda_N)]}}{\exp{[w_a/(2\lambda_N^*)]}}\simeq \frac{w_A}{2\lambda_N^*}\frac{\exp{[w_a/(2\lambda_N)]}}{\exp{[w_a/(2\lambda_N^*)]}}$.

\vspace{0.1in}

\textit{vi}) The use of the point-contact model leads to a systematic underestimation of $\lambda_{A}$ for $w_A>\lambda_N^*$ towards a smaller apparent value $\lambda_{A}^*$. The ratio between apparent and real values of $SDL$ in $A$ approaches the ratio of $\eta$ according to $\frac{\lambda_{A^*}}{\lambda_A}=\frac{w_A}{2\lambda_N^*}\frac{\exp{[w_a/(2\lambda_N)]}}{\exp{[w_a/(2\lambda_N^*)]}}$.

\vspace{0.1in}

\begin{figure}[H]
\begin{centering}
\includegraphics[width=8cm]{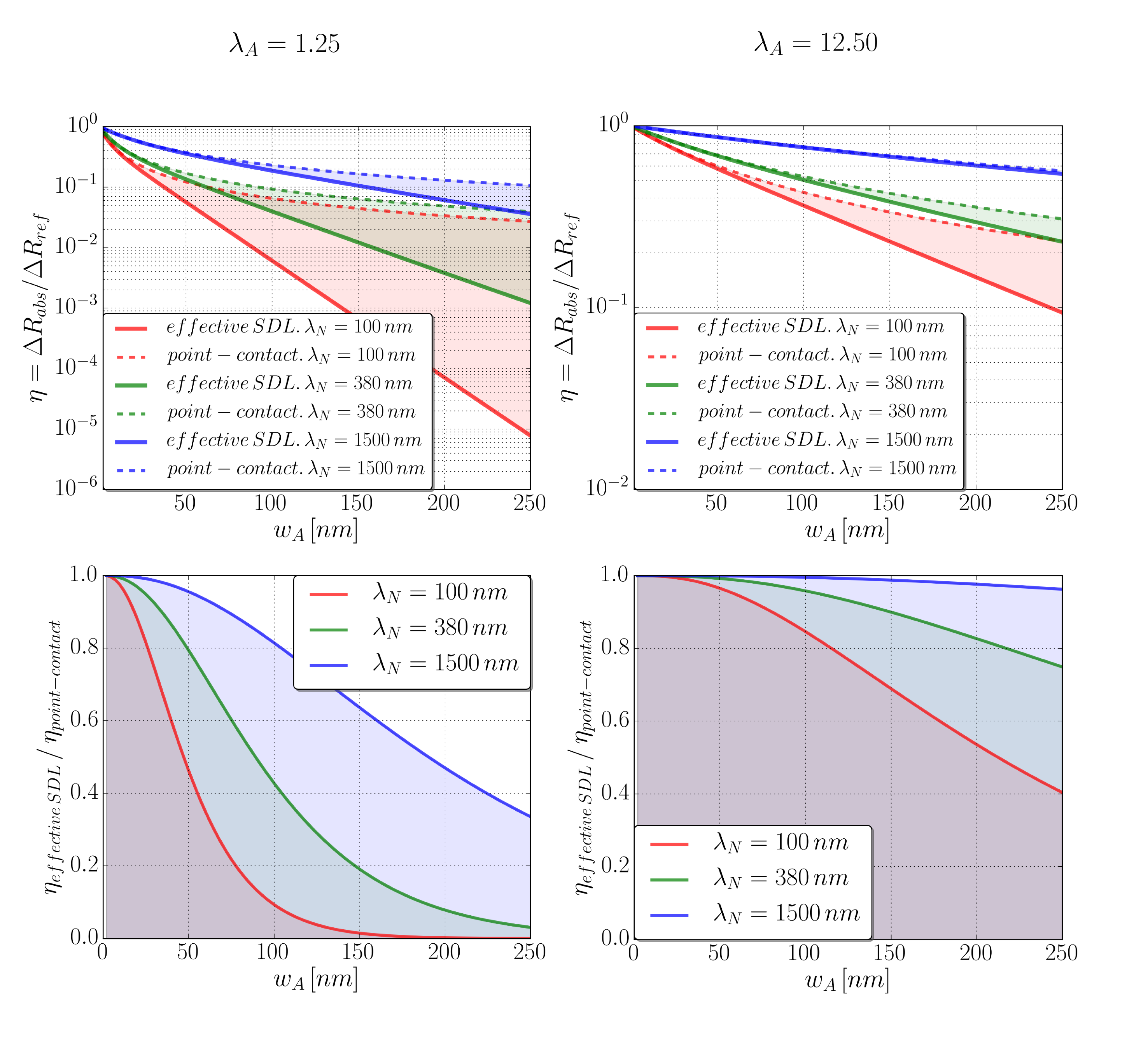}
\par\end{centering}
\caption{\textit{\label{fig:FIG5}Numerical calculations of the normalized spin signal $\eta$ (top) and its ratio for the point-contact and the extended models as a function of $AuW$ width $w_{A}$ for the case of $\lambda_{A}=1.25\,nm$ (left) and $\lambda_{A}=12.5\,nm$ (right) with variable }$\lambda_{N}$\textit{.}}
\end{figure}

\textit{vii}) At constant resistivity $\rho_N$, the larger $\lambda_N$ is the larger is the effect of the renormalization by the effective $SDL$ model.
From Eq.[2], the effective $SDL$ $\lambda_{N}^*$ saturates to a constant value for large values of $\lambda_N$ above $\sqrt{\frac{\rho_A}{\rho_N} \lambda_A t_N}$ which makes the renormalization insensitive to $\lambda_N$ in that region. Accordingly, the error becomes independent of $\lambda_N$ above $\sqrt{\frac{\rho_A}{\rho_N} \lambda_A t_N}$. The top part of Fig.~\ref{fig:FIG5} displays the normalized spin-resistance $\eta=\frac{\Delta R_{abs}}{\Delta R_{ref}}$ \textit{vs.} contact width $w_{A}$ for respective point-contact and effective $SDL$ (dashed and continues lines respectively) whereas the bottom part displays the ratio of respective $\eta$. Fig.~\ref{fig:FIG5} displays some of calculations for $\eta$, represented by different colors which correspond to three different $\lambda_{N}$.

\vspace{0.1in}

\textit{viii}) In the fitting procedures, one may consider the case of a constant spin-resistance $R_N$ and variable $\lambda_N$ (\textit{e.~g.} by varying continuously $\rho_N$ oppositely to $\lambda_N$ when the latter is changed). In that case the effective $SDL$ $\lambda_{N}^*$ follows the variation of $\lambda_N$ from Eq.~[2]. Here, the smaller $\lambda_N$ is the larger is the effect of the renormalization to perform. Calculations were made for the case of a short $\lambda_{A}=1.25\,nm$ {[}Fig. \ref{fig:FIG5}{]}. More generally, the differences between the point-contact and the effective $SDL$ models depend strongly on several parameters and a general description of their limits cannot be simply given, as illustrated in Fig.~\ref{fig:FIG5} in the case of a short $\lambda_{A}$ ($1.25\,nm$ in the present case even in the case of a long $\lambda_{N}$ (up to $1500\,nm$)).

\vspace{0.1in}

\textit{ix}) One can give an exact literal expression of the spin-signal in the case of the extended spin-sink (or the effective $SDL$ model) placed in between the two ferromagnetic injectors $F$ (see Appendix).

\section{Conclusions}

In conclusion, we emphasize that the extraction of the $SDL$ should be carefully addressed, particularly when the spin accumulation and spin-current polarization profiles along the width of a spin absorber cannot be considered uniform any longer for large spin-sink sizes and especially for small $\lambda_{A}$. We have addressed this issue by developing adapted models taking into account this variation. The validity of this approach was experimentally proven by studying the absorption of the spin accumulation as a function of the $AuW_{13.8\%}$ spin-absorber width in non-local spin valve experiments. We have demonstrated that the geometrical effects have to be considered with caution in the analysis of the spin-signals in non-local and local geometry for spin-sink experiments and we have refined the $1D$ model accordingly. Finally, we have presented a unique method which is well adapted to precisely determine short spin diffusion lengths by studying the width dependence of the spin current absorption in lateral spin valves with the inserted spin-absorber.

\begin{acknowledgments}
The samples were fabricated in the Plateforme \textit{Technologie Avancée} in Grenoble, for which we acknowledge the support of the Renatec network. This work was supported by the ANR simi 10 SOspin project.
\end{acknowledgments}

\section{Appendix}

The mathematical expressions for the spin signal for the 3 different systems: the reference $LSV$ without spin-absorber, the $LSV$ with absorber within the point-contact scheme and the $LSV$ with absorber within the effective $SDL$ approach are expressed in the framework of the \textit{Spin-Resistor} model like developed in details in Ref.~\citep{SaveroTorres_Xresistor_2015}:

\begin{itemize}
\item \textbf{The spin signal of the reference lateral structure} (without spin-absorber) can be written in a matrix form as follows:
\end{itemize}

\begin{center}
	$\begin{array}{c}
	\Delta R_{ref}=2\left(\begin{array}{c}
	0\\
	{P_{F}}
	\end{array}\right).\left(\begin{array}{cc}
	p & -q\\
	-q & p
	\end{array}\right)^{-1}\left(\begin{array}{c}
	{P_{F}}\\
	0
	\end{array}\right)\\
	\\
	=2q{P_{F}}^{2}/\left(p^{2}-q^{2}\right)
	\end{array}$
	\par\end{center}

with $p=\frac{1}{R_{F}^{*}}+\frac{1}{R_{N}}+\frac{1}{R_{N}\tanh(L/\lambda_{N})}$ and $q=\frac{1}{R_{N}\sinh(L/\lambda_{N})}$. It corresponds to the well known expression given in Ref.~\citep{Jedema_PRB_2003,Kimura_PRB_2005}.

\begin{itemize}
\item \textbf{For the device with an inserted spin-absorber} (considering the point-contact assumption for the spin-absorber), one can define the spin-signal as:
\end{itemize}

\begin{center}
	$\begin{array}{c}
	\Delta R_{abs}^{point-contact}=\\2\left(\begin{array}{c}
	0\\
	0\\
   {P_{F}}
	\end{array}\right).\left(\begin{array}{ccc}
	f & -h & 0\\
	-h & g & -h\\
	0 & -h & f
	\end{array}\right)^{-1}\left(\begin{array}{c}
	{P_{F}}\\
	0\\
	0
	\end{array}\right)\\
	\\
	=2h^{2}{P_{F}}^{2}/(f^{2}g-2fh^{2})
	\end{array}$
	\par\end{center}

with $f=\frac{1}{R_{N}}+\frac{1}{R_{F}^{*}}+\frac{1}{R_{N}\tanh(\frac{L}{2\lambda_{N}})}$, $g=\frac{\tanh(\frac{t_{A}}{\lambda_{A}})}{R_{A}}+\frac{2}{R_{N}\tanh(\frac{L}{2\lambda_{N}})}$,
$h=\frac{1}{R_{N}\sinh(\frac{L}{2\lambda_{N}})}$, in the limit of a short spin diffusion length in the ferromagnets $F$ and the spin-absorber \footnote{Note that the commonly used formula proposed by Kimura \textit{et al.}~\citep{Kimura_PRB_2005}, represented in the resistor approach formalism would require the use of the following coefficients: $f=\frac{1}{R_{N}}+\frac{2}{R_{F}^{*}}+\frac{1}{R_{N}\tanh(\frac{L}{2\lambda_{N}})}$,
$g=\frac{2}{R_{A}}+\frac{2}{R_{N}\tanh(\frac{L}{2\lambda_{N}})}$, $h=\frac{1}{R_{N}\sinh(\frac{L}{2\lambda_{N}})}$. The factor 2 in the fractions of $R_{F}^{*}$ and $R_{A}$ appears because Kimura \textit{et al.} have considered the 2 branches of $F$ and $A$ wires. However, since the spin diffusion lengths of these materials are short, their spin-resistances should be counted only once (only 1 branch of $F$ and $A$ wires)}.
\begin{itemize}
\item \textbf{For the device with an inserted spin-sink (considering the extended effective $SDL$ model)} one gets:
\end{itemize}

\begin{center}
	$\begin{array}{c}
	\Delta R_{abs}^{effective\,SDL}=\\2\left(\begin{array}{c}
	0\\
	0\\
	0\\
	{P_{F}}
	\end{array}\right).\left(\begin{array}{cccc}
	a & -c & 0 & 0\\
	-c & b & -d & 0\\
	0 & -d & b & -c\\
	0 & 0 & -c & a
	\end{array}\right)^{-1}\left(\begin{array}{c}
	{P_{F}}\\
	0\\
	0\\
	0
	\end{array}\right)\\
	\\
	=2c^{2}d{P_{F}}^{2}/((c^{2}-ab)^{2}-(ad)^{2})
	\end{array}$
	\par\end{center}

where the required coefficients are given by:

\begin{center}
$\begin{array}{c}
a=\frac{1}{R_{N}}+\frac{1}{R_{F}^{*}}+\frac{1}{R_{N}\tanh(\frac{L-w_{A}}{2\lambda_{N}})}\\
b=\frac{1}{R_{N}\tanh(\frac{L-w_{A}}{2\lambda_{N}})}+\frac{1}{R_{A}^*\tanh(\frac{w_{A}}{\lambda_{N}^*})}\\
c=\frac{1}{R_{N}\sinh(\frac{L-w_{A}}{2\lambda_{N}})}\\
d=\frac{1}{R_{A}^*\sinh(\frac{w_{A}}{\lambda_{N}^*})}
\end{array}$
\par\end{center}

The spin resistances used above are defined as:

$R_{N}=\frac{\rho_{N}\lambda_{N}}{w_{N}t_{N}}$, $R_{F}^{*}=\frac{\rho_{F}\lambda_{F}}{(1-P_F^{2})w_{F}w_{N}}$,
$R_{A}=\frac{\rho_{A}\lambda_{A}}{w_{A}w_{N}}$, $R_{A}^*=\frac{\rho_{N}\lambda_{N}^*}{t_{N}w_{N}}$, where the effective spin diffusion length is:

\begin{equation}
\lambda_{N}^*=\frac{\lambda_{N}}{\sqrt{1+\frac{\rho_{N}}{\rho_{A}}\frac{\lambda_{N}^{2}}{\lambda_{A}t_{N}}\tanh(\frac{t_{A}}{\lambda_{A}})}}=\frac{\lambda_N}{\sqrt{1+\frac{\mathcal{R}_N}{\mathcal{R}_A}\tanh(\frac{t_A}{\lambda_A}) }}
\end{equation}

\end{document}